\pdfoutput=1
\documentclass[floatfix,altaffilletter,preprintnumbers,superscriptaddress,tightenlines,showpacs,showkeys,nofootinbib]{revtex4-1}

\usepackage[dvips]{graphicx}
\usepackage{array,amsmath,amsthm,feynmf,mathtools}
\usepackage{amssymb}
\usepackage{braket}
\usepackage{breakurl}
\usepackage{color}
\usepackage{enumitem}
\usepackage{epsfig}
\usepackage{graphicx}
\usepackage{hyperref}
\usepackage{mathtools}
\usepackage{siunitx}
\usepackage{slashed}
\usepackage{xcolor}
\usepackage[normalem]{ulem}

\usepackage{lmodern}
\usepackage{tikz}

\usepackage[T1]{fontenc}

\makeatletter

\begin{document}                                                                                                                                                                                                                                                                                                                                                                                                                                                                                                                                                                                                                                                                    
\begin{flushright}
	CPHT-RR077.122023
\end{flushright}

\title{Prediction of the bubble wall velocity for a large jump in degrees of freedom}

\author{Mikel Sanchez-Garitaonandia}
\affiliation{CPHT, CNRS, \'Ecole polytechnique, Institut Polytechnique de Paris, 91120 Palaiseau, France}

\author{Jorinde van de Vis}
\affiliation{Institute for Theoretical Physics, Utrecht University, Princetonplein 5, 3584 CC Utrecht, The Netherlands}
\affiliation{Instituut-Lorentz for Theoretical Physics, Leiden University, Niels Bohrweg 2, 2333 CA Leiden, the Netherlands}

\begin{abstract}
The bubble expansion velocity is an important parameter in the prediction of gravitational waves from first order phase transitions. This parameter is difficult to compute, especially in phase transitions in strongly coupled theories.
In this work, we present a method to estimate the wall velocity for phase transitions with a large enthalpy jump, valid for weakly and strongly coupled theories. We find that detonations are disfavored in this limit, but wall velocities are not necessarily small.
 We also investigate the effect of two other features in the equation of state: non-conformal sound speeds and a limited range of temperatures for which the phases exist. We find that the former can increase the wall velocity for a given nucleation temperature, and the latter can restrict the wall velocities to small values. To test our approach, we use holographic phase transitions, which typically display these three features. We find excellent agreement with numerically obtained values of the wall velocity. We also demonstrate that the implications for gravitational waves can be significant.

\end{abstract}

\maketitle

\section{Introduction}
\label{sec:Intro}
Many models for particle physics beyond the Standard Model (BSM) predict that one or more first order phase transitions (PTs) might have occurred in the history of the universe. These phase transitions might have played a role in the generation of the matter-antimatter asymmetry \cite{Kuzmin:1985mm, Shaposhnikov:1986jp, Shaposhnikov:1987tw, Cohen:1990py, Cohen:1993nk,Katz:2016adq,  Baldes:2021vyz, Azatov:2021irb} or the production of dark matter \cite{Baker:2019ndr, Chway:2019kft, Azatov:2021ifm, Baldes:2022oev}.
First order PTs can also source a gravitational wave (GW) signal when the released vacuum energy gets converted into sound waves, gradient energy in the bubble walls and/or turbulence~\cite{Witten:1984rs, Kosowsky:1991ua, Kosowsky:1992rz, Kamionkowski:1993fg}. 
Depending on the strength and temperature of these transitions, the signals could be observable with the next generation of GW telescopes~\cite{Grojean:2006bp, Caprini:2018mtu,Caprini:2019egz}. 
See e.g. \cite{Weir:2017wfa, Caprini:2019egz, Hindmarsh:2020hop, Athron:2023xlk} for reviews.

Using GWs to learn about BSM physics requires accurate predictions of the GW spectrum. Analytical arguments and hydrodynamic simulations
have resulted in a predicted GW spectrum 
as a broken power-law~\cite{Hindmarsh:2015qta, Hindmarsh:2017gnf, Caprini:2019egz,Hindmarsh:2016lnk, Hindmarsh:2019phv, Jinno:2020eqg, Jinno:2022mie, Cai:2023guc, RoperPol:2023dzg}.
The underlying assumption in~\cite{Caprini:2019egz, Hindmarsh:2016lnk, Hindmarsh:2019phv, Jinno:2020eqg, Jinno:2022mie, RoperPol:2023dzg}, which was checked in~\cite{Hindmarsh:2015qta, Hindmarsh:2017gnf}, 
is that the amplitude of the signal can be predicted by the nucleation temperature $T_n$ and rate $\beta$, wall velocity $\xi_w$ and sound speed $c_s$,
as well as the so-called energy budget, that can be computed from the hydrodynamics of
a single bubble.\footnote{
Strictly speaking, the GW signal should be computed at the percolation temperature $T_p$. Often one can approximate $T_p \sim T_n$, but this assumption can not always be made \cite{Ellis:2018mja, Athron:2022mmm, Athron:2023rfq}. 
}
 In most cases, the energy budget can be estimated from the phase transition strength $\alpha$, $\xi_w$~\cite{Espinosa:2010hh}, and $c_s$~\cite{Giese:2020rtr, Giese:2020znk}. 
The challenge of predicting the GW spectrum is thus reduced to a computation of the thermal parameters and the hydrodynamics. 

The wall velocity strongly affects the strength and shape of the GW spectrum~\cite{Espinosa:2010hh,Caprini:2019egz}, and it also enters in computations of the baryon asymmetry \cite{Bodeker:2004ws, Fromme:2006cm, DeVries:2018aul, Cline:2020jre, Dorsch:2021ubz, Cline:2021dkf} or dark matter abundance \cite{Baker:2019ndr, Chway:2019kft, Azatov:2021ifm, Baldes:2022oev}.
It is however challenging to compute for a given theory 
(see however~\cite{Moore:1995si, Dorsch:2018pat,Lewicki:2021pgr, Laurent:2022jrs, Jiang:2022btc} for some explicit computations) and 
is therefore often treated as a free parameter, or simply set to $\xi_w \to 1$. This results in a significant uncertainty, so better estimates are necessary.
Unfortunately, the difficulty of computing the wall velocity is even greater in strongly coupled theories~\cite{Schwaller:2015tja, Bruggisser:2018mrt, Huang:2020crf, Halverson:2020xpg, Yang:2022ghc}, 
where a quasiparticle interpretation underlying the computations of~\cite{Moore:1995si, Dorsch:2018pat,Lewicki:2021pgr, Laurent:2022jrs, Jiang:2022btc} may not be available. 
In this work, we will 
provide a way to determine $\xi_w$ that can be applied to both weakly and strongly coupled theories when they have a large jump
in the number of degrees of freedom between the phases -- we will refer to this as a large enthalpy jump. 
We will take a holographic equation of state (EoS) as an explicit example,
which can be used to understand aspects of strongly coupled PTs~\cite{Nitti:2008za,Nitti:2009jd}. Holographic models represent a useful playground 
where we can obtain insight into strongly coupled cosmological PTs~\cite{Bigazzi:2020phm, Ares:2020lbt,Bigazzi:2020avc,Ares:2021ntv, Zhu:2021vkj, Morgante:2022zvc,Chen:2022cgj} and possibly into PTs in neutron star mergers~\cite{Casalderrey-Solana:2022rrn}.
It has already been shown that $\xi_w$ can be obtained in a holographic PT~\cite{Bea:2021,Bea:2022,Bigazzi:2021ucw,Janik:2022wsx}, 
and we can use these results to test our approach. 

Holographic theories are characterized by a gravitational description that is dual to a strongly coupled gauge theory with a large number $N$ of degrees of freedom.
They naturally exhibit three distinctive features (see e.g.~\cite{Bea:2018whf,Elander:2020rgv}):
\begin{itemize}
	\item{A large jump in enthalpy between the high-temperature and low-temperature phase,}
	\item{A limited range of temperature for which the phases exist,}
	\item{Strong deviation in the sound speed from the conformal value $c_s^2 = 1/3$.}
\end{itemize} 
We will see that these features can strongly modify the hydrodynamic predictions compared to the often-made assumptions that $c_s^2 \sim 1/3$ and that the 
temperature can take any value.\footnote{Sometimes this assumption is implicit, by using an approximation for the energy budget 
of~\cite{Espinosa:2010hh, Giese:2020rtr, Giese:2020znk}. The underlying models of~\cite{Espinosa:2010hh, Giese:2020rtr, Giese:2020znk}
exist for arbitrary temperatures,
and the corresponding solutions could probe temperatures unrealistically far away from $T_n$.
}
We also stress that these features are not unique to holographic PTs, and that our results apply to non-holographic models as well.

The main result of this work is a demonstration that the wall velocity for models with a large enthalpy jump follows directly
from the EoS and the nucleation temperature, without further details of the plasma. Some steps in this direction were already taken in~\cite{Janik:2022wsx}, where a formula was proposed to obtain the wall speed for planar bubbles. Our results are valid for spherical bubbles and we will compare our results for planar bubbles with the findings there. Additionally, detonations are not realizable in the infinite enthalpy jump limit.
If the allowed temperature range is limited, we find that the resulting wall velocity is rather small, favoring deflagrations as observed in~\cite{Bea:2021,Bea:2022,Janik:2022wsx}, and likely excluding detonations.
Finally, we will see that quantitative results can get strongly affected by a non-conformal sound speed. GW predictions for models with a large enthalpy jump get significant corrections compared to the 
`vanilla' assumption where $\xi_w$ is a free parameter and $c_s^2 = 1/3$.

\section{Hydrodynamic and thermodynamic description of bubble expansion}
\subsection{Hydrodynamics}\label{sec:hydro}
Finding the kinetic energy in the fluid requires solving the hydrodynamic equations of a single expanding bubble. 
We will summarize the approach here, and further details can be found in~\cite{landau1987fluid, Kurki-Suonio:1995rrv, Kamionkowski:1993fg,Espinosa:2010hh}. 
The hydrodynamic equations follow from the energy-momentum tensor of a perfect fluid, 
\begin{equation}
T^{\mu\nu} = w u^{\mu}u^{\nu}-pg^{\mu\nu},
\end{equation}
where $w = T dp/dT$ is the enthalpy. 
$g^{\mu\nu}$ denotes the metric, which is assumed to be the Minkowski metric, and $u^\mu = \gamma(1, \vec v)$ denotes the fluid velocity with $\gamma = 1/\sqrt{1- \vec v^2}$.
The hydrodynamic equations for $u^\mu$ and $w$ are obtained by projecting $\partial_\mu T^{\mu\nu} = 0$ in the direction parallel and perpendicular to the fluid flow:
 \begin{equation}
\begin{aligned}
2\frac{v}{\xi} & = \gamma^2(1-v\xi)\left[\frac{\mu^2(\xi,v)}{c_s^2}-1\right]\partial_{\xi}v,\\
\partial_{\xi} w & = w\left(1+\frac{1}{c_s^2}\right)\gamma^2\mu(\xi,v)\partial_{\xi}v,
\end{aligned}
\label{eq:self_similar}
\end{equation} 
where $\xi=r/t$ (with $r$ the radial distance to the center of the bubble and $t$ the time since nucleation), $v$ the fluid velocity in radial coordinates
and $\mu$ is the Lorentz-boosted velocity, $\mu(\xi,v) = (\xi-v)/(1-\xi v)$.
The speed of sound follows from $p$ via 
\begin{equation}
  c_s^2 = \frac{dp/dT}{de/dT}, \qquad e = p \frac{dp}{dT} - p.
\end{equation}
The only regions where this hydrodynamic description fails are the bubble walls and shocks. Nevertheless, given their small size compared to the bubble, 
they can be replaced by discontinuities across which we impose matching conditions, that are
obtained by integration $\partial_\mu T^{\mu\nu} = 0$ from right behind the wall/shock to right in front of the wall/shock, yielding 
\begin{equation}
\begin{aligned}\label{eq:matching0}
  w_+ v_+^2 \gamma_+^2 + p_+ &= w_-v_-^2\gamma_-^2 + p_-, \\
   w_+ v_+\gamma_+^2 &= w_- v_- \gamma_-^2,
\end{aligned}
\end{equation}
where the $+,(-)$ label denotes quantities evaluated right in front of (behind) the discontinuity.
After some algebraic manipulations, we obtain
\begin{equation}
  v_+v_- = \frac{p_+-p_-}{e_+-e_-},\quad	\quad	\frac{v_+}{v_-} = \frac{e_-+p_+}{e_++p_-},
  \label{eq:matching}
\end{equation}
for the matching conditions.
Note that the velocities are defined with respect to the frame of the wall/shock.

By solving eqs.~\eqref{eq:self_similar} and imposing the matching conditions eqs.~\eqref{eq:matching} at the wall and possible shock we can solve for the whole flow as a function of two input parameters, e.g. $\xi_w$ and $T_n$ (or, similarly, $\alpha_n$).
The energy budget $K$ is determined from the ratio of the fluid kinetic energy to the enthalpy of the high-temperature phase at $T_n$.
For each $T_n$, there are three type of solutions depending on the value of the wall speed with respect to the sound speed (see, e.g.~\cite{Espinosa:2010hh}):
\begin{itemize}
  \item{Deflagrations: $\xi_w< c_{s,L}$. In this case $v_-=\xi_w$ and $v_+ < c_{s,H}(T_+)$.  }
  \item{Hybrids: $c_{s,L}< \xi_w < v_J$, with $v_J$ the Jouguet velocity (see e.g.~\cite{Kurki-Suonio:1995rrv,Espinosa:2010hh,Ai:2023see}). We now have $v_- = c_{s,L}$ and $v_+ <= c_{s,H}$.}
  \item {Detonations: $\xi_w > v_J$, with $v_- > c_{s,L}$ and $v_+ =\xi_w$.}
\end{itemize}
Here and in the following we use the subscript $H$ ($L$) for quantities defined in the high- (low-) enthalpy phase.

\subsection{Equations of state}
\begin{figure}[htbp]
  \begin{center}
  \includegraphics[width=0.45\textwidth]{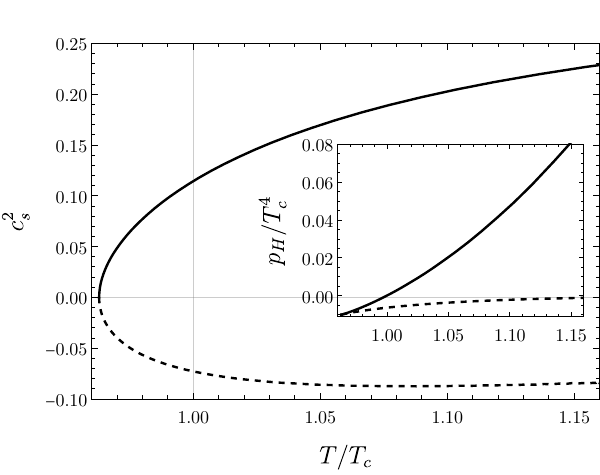}
  \caption{Speed of sound squared and pressure (inset) of the high enthalpy phase. 
  The dashed lines correspond to states 
  that are thermodynamically unstable. 
  }
  \label{fig:high_eos}
  \end{center}
\end{figure}

We will consider three equations of state to investigate the effects of large enthalpy jumps, temperature limitations and $c_s^2 \neq 1/3$ 
on the hydrodynamic solutions. 
Let us parametrize the large enthalpy jump by some large number $N$, and assume that low-enthalpy phase quantities are suppressed by\footnote{
 The inspiration for this suppression factor comes from confining holographic and large-$N$ gauge theories, where the high-enthalpy phase has $N^2$ degrees of freedom fand the low-enthalpy phase $\mathcal O(1)$. However, we do not take any other assumptions from holography, so the results we obtain are general for models with a large enthalpy jump.
    }
$1/N^2$.

\paragraph*{Dark SU(N) model}
We consider the model of~\cite{Nitti:2008za,Nitti:2009jd}, which is based on a holographic description of a confinement/deconfinement PT. 
The thermodynamic equation of state can be obtained following the approach described in~\cite{Nitti:2008za}, 
which we will not repeat here, but we 
demonstrate the pressure and sound speed of the high-enthalpy phase in fig.~\ref{fig:high_eos}.
We see that the sound speed deviates strongly from $c_s^2 = 1/3$. At the nucleation temperature $T_n = 0.993 T_c$~\cite{Morgante:2022zvc},
$c_s^2 = 0.103$. Note that the model only describes the high-enthalpy phase, which ceases to exist at $T<T_{\rm min, SU(N)}$.
Even though we leave $N$ as a free parameter, the model implicitly assumes that $N$ is large, as
contributions that are smaller than $\mathcal O(N^2)$ are neglected.
For the low-enthalpy phase, we will assume the equation of state proposed by~\cite{Leitao:2014pda} (although we will find that nothing depends on this 
for large $N$)
\begin{equation}\label{eq:templateLow}
  p_L = \frac{1}{N^2}\frac{T_c^4}{\nu -1}  \left( \frac{T}{T_c}\right)^\nu,
\end{equation}
where the sound speed is a constant set by $\nu = 1 + 1/c_{s,L}^2$, which parameterizes the unknown equation of state
of the low-enthalpy phase.

\paragraph*{Template model}
We use the model of~\cite{Leitao:2014pda} to describe models with constant, but non-conformal sound speeds,
\begin{equation}
\begin{aligned}
  p_H &= \frac{a_H T_c^4}{\mu -1} \left(\frac{T}{T_c}\right)^\mu - \epsilon, \\ p_L &= \frac{1}{N^2}\frac{a_H T_c^4}{\nu -1}  \left( \frac{T}{T_c}\right)^\nu,
\end{aligned}
\end{equation}
where and $\mu = 1 + 1/c_{s,H}^2$ and $a_H$ parameterizes the number of degrees of freedom of the high-enthalpy phase and $\epsilon$ is a temperature-independent constant,
which parameterizes the energy difference between the two phases.
The virtue of the template model is that it allows us to study the effect of the non-conformal sound speed
and large enthalpy jump, without the limited temperature range. The often-used bag equation of state is a special case of the template model
with $\mu=\nu = 4$.

In all cases, we assume that the low-enthalpy phase temperature cannot grow as large as to undo the $1/N^2$ suppression, i.e. we assume that $T_L\ll T_c N^{2/\nu}$.

\paragraph*{Strongly coupled holographic model}
We will briefly discuss the model of~\cite{Bea:2022} in sec.~\ref{sec:compare}, a holographic description of a strongly coupled phase transition with $N\sim 3$.
(see~\cite{Bea:2022} for details).
The bubble wall velocity of this model was determined in~\cite{Bea:2021}, and we will compare it to our estimate.

\section{Hydrodynamics with a large enthalpy jump}
\subsection{Matching conditions in the large $N$ limit}
Let us consider the matching conditions of eq.(\ref{eq:matching}) in the limit of $N \rightarrow \infty$.
Given the suppression that we assumed for the low-enthalpy phase, the 
following ratios hold, 
\begin{equation}
\frac{w_-}{w_+} \sim \frac{s_-}{s_+} \sim \frac{e_-}{e_+} \sim \frac{1}{N^2},
\end{equation}
where $s = w/T$.
We have assumed the previously mentioned bound on the low-enthalpy phase temperature. 

The relation between the pressures of both phases is more complicated. By the 
definition of $T_c$, $p_H(T_c)=p_L(T_c)$, which means that $p_-/p_+$ does not 
need to be small. Indeed, 
\begin{equation}
\begin{aligned}\label{eq:cases}
\frac{p_-}{p_+}\sim \frac{1}{N^2}\quad &\mathrm{if} \quad \frac{|T_+-T_c|}{T_c} \gg \frac{1}{N^2} , \qquad \textbf{Case 1,}\\
\frac{p_-}{p_+}\sim 1 \quad &\mathrm{if}\quad \frac{|T_+-T_c|}{T_c} \lesssim \frac{1}{N^2}, \qquad \textbf{Case 2.}
\end{aligned}
\end{equation}
We will now study the matching conditions for these two different cases. 

\paragraph*{Case 1}
Let us first investigate eq.~\eqref{eq:matching} with $p_-/p_+ \sim 1/N^2$,
\begin{equation}
v_+v_-  \sim  \frac{p_+}{e_+}\left(1+\mathcal{O}\left(\frac{1}{N^2}\right)\right)\sim \frac{v_+}{v_-},
\end{equation}
with solution\footnote{
Note that we use the sign conventions of~\cite{Espinosa:2010hh}, where all velocities are positive.
} $v_- = 1-\mathcal{O}(1/N^2)$, which corresponds to a 
detonation, as we have seen in sec.~\ref{sec:hydro}.
In a detonation, $v_+>v_-$, so eq.~\eqref{eq:matching0} implies that $w_->w_+$, which is not possible in the large $N$ limit.\footnote{
 Even without taking the large $N$ limit, detonations can be excluded in certain theories with a limited temperature range as having $w_->w_+$ may not be possible.}

\paragraph*{Case 2}
If, $p_-/p_+ \sim 1$, then,
\begin{equation}
v_+v_-  =  \mathcal{O}\left(\frac{1}{N^2}\right), \qquad 
\frac{v_+}{v_-}  = \mathcal{O}\left(\frac{1}{N^2}\right),
\end{equation}
implying that $v_+=\mathcal{O}(1/N^2)$, corresponding to a deflagration or a hybrid. In combination with eq.~\eqref{eq:cases} this leads to the following conditions for a large enthalpy jump bubble expansion,
\begin{equation}
  T_+=T_c\quad\mathrm{and}\quad v_+=0.
  \label{eq:Large-N-aprox}
  \end{equation}
The condition on $v_+$ was already stated in~\cite{Janik:2022wsx} and it was also pointed out there that the condition on $T_+$ was a good approximation for the simulations presented in~\cite{Bea:2021,Janik:2022wsx}.

Let us point out that the matching relation just obtained does not restrict us to slow walls, as the Jouguet velocity,
which is the transition from hybrids to detonations can become arbitrarily close to unity for a large amount of supercooling (see e.g.~\cite{Espinosa:2010hh}).

We conclude that large enthalpy jump PTs lead to deflagrations or hybrids characterized by the condition~\eqref{eq:Large-N-aprox}  while detonations are excluded.

\subsection{Hydrodynamic solutions}
We expect the matching relations of eq.~\eqref{eq:Large-N-aprox} to hold in any theory with a large
enthalpy jump, but in order to use them to find the relation between $\xi_w$ and $T_n$ 
requires a solution of the hydrodynamic equations in the shock wave. This requires an EoS, and the relation between $\xi_w$ and $T_n$ will be numerical. 

In this section, we solve the hydrodynamics for the dark SU(N) and the template model (with $c_s^2 = 1/3$ and $c_s^2 = 0.103$ -- the same value as in the dark SU(N)), for increasing $N$. 

\begin{figure}[htbp]
  \begin{center}
  \includegraphics[width=0.45\textwidth]{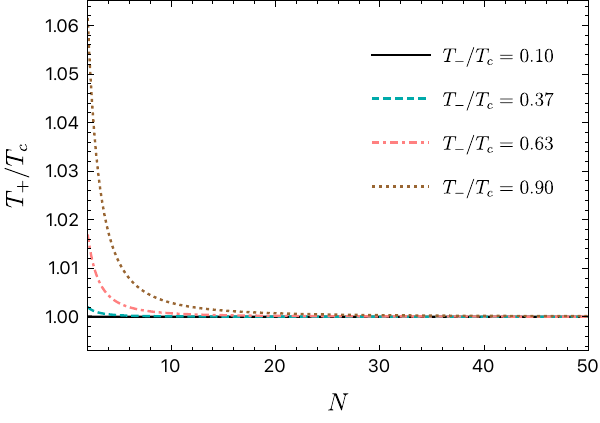}
  \includegraphics[width=0.45\textwidth]{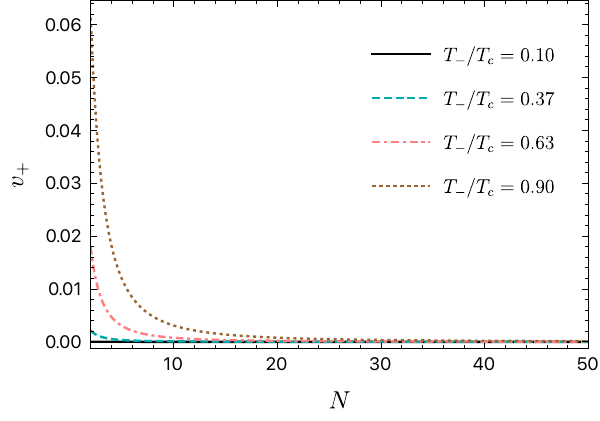}    
  \caption{$T_+/T_c$ (Left) and $v_+$ (Right) as a function of $N$ for different choices of $T_-$. $T_+$ tends to $T_c$ and $v_+$ to zero as $N$ increases regardless of $T_-$.}
  \label{fig:large-N-condition-check}
  \end{center}
\end{figure}

Let us focus first on deflagrations and hybrids. We demonstrate in the left panel of fig.~\ref{fig:large-N-condition-check} that $T_+$ indeed approaches $T_c$, independent of the choice of $T_-$. The slower converge for larger $T_-$ is to be expected as we require larger $N$ to suppress the enthalpy at higher temperatures. Similarly, $v_+$ approaches 0 as $N$ increases, independently of  the value of $T_-$, as shown in the right panel of fig.~\ref{fig:large-N-condition-check}.
We used the template model with $\mu,\nu = 4$, but have confirmed explicitly that the convergence also occurs for other sound speeds and the dark SU(N) model.

\begin{figure}[htbp]
  \begin{center}
  \includegraphics[width=0.45\textwidth]{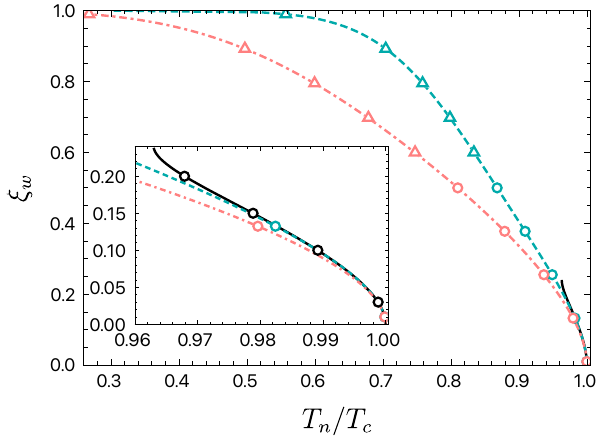}    
  \caption{Wall speed as a function of the nucleation temperature in the dark SU(N) model (black), template model with $c_s^2 = 1/3$ (pink dot-dashed) and with $c_s^2 = 0.103$ (cyan dashed). Circles correspond to deflagrations and triangles to hybrids.}
  \label{fig:large-N-speed}
  \end{center}
\end{figure}

The lines of fig.~\ref{fig:large-N-speed} demonstrate the relation $\xi_w(T_n)$ obtained from
solving the hydrodynamic equations with the matching conditions of eq.~\eqref{eq:Large-N-aprox}. In this limit, the 
solution is obtained without any reference to the fluid behind the wall. The
dots demonstrate the result for $N=30$ (with $\nu=4$ and $T_-/T_c=0.5$) and we see that they agree extremely well. 
We observe that for the holographic model, only solutions with $\xi_w \lesssim 0.25$ exist. The reason
becomes immediately clear -- the minimum temperature of the high-enthalpy phase prevents further
supercooling. From the results of the template model, we see that when such a minimum temperature does 
not exist, there is -- in principle -- no limit on the amount of supercooling and the wall velocity can
grow arbitrarily large. This is a very important observation, as it suggests that the low wall
velocities found in real-time holographic simulations 
\cite{Janik:2022wsx,Bea:2021,Bea:2022} are not a result of the strongly coupled nature of these theories,
but rather of the impossibility of strong supercooling.

Another interesting observation is that the sound speed strongly affects the relation between
$T_n$ and $\xi_w$. For models with $c_s^2 < 1/3$, the typical velocity is larger than for $c_s^2 = 1/3$, and
a smaller amount of supercooling is required for a fast wall, in agreement with~\cite{Janik:2022wsx}.

Finally, we checked that detonations get excluded in the large enthalpy jump limit for the template model. Given a fixed $\xi_w$ and $T_n$, eq.~\eqref{eq:matching} implies that $p_-$ and $e_-$ are independent of $N$, meaning that the temperature scales as $T_-\propto T_c N^{2/\nu}$ for detonations. But this corresponds precisely to the range of temperatures that we excluded in the large $N$ limit because it undoes the $1/N^2$ suppression.

\subsection{Comparison with the strongly coupled holographic model}\label{sec:compare}
\begin{figure}[htbp]
  \begin{center}
  \includegraphics[width= 0.45\linewidth]{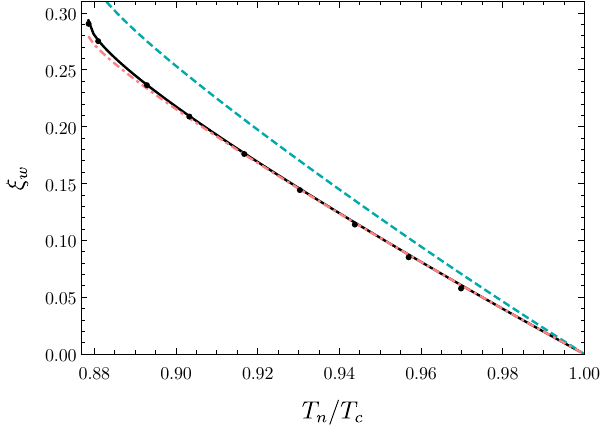}
  \caption{Wall speed from the simulations in~\citep{Bea:2021} (dots), and the result using the condition \eqref{eq:Large-N-aprox} (solid black), local thermal equilibrium (dashed cyan) and the simple wave formula \eqref{eq:simple-wave} (dot-dashed pink).}
  \label{fig:apply-to-Mikel-linear}
  \end{center}
\end{figure}

Let us now put our result to the test, by comparing to the wall velocity obtained in an actual simulation. Fig.~\ref{fig:apply-to-Mikel-linear} demonstrates a comparison to the values of $\xi_w$ obtained in the strongly coupled holographic model in~\cite{Bea:2021} for planar bubbles. In this case, $\partial_\mu T^{\mu\nu} = 0$ reduces to $\partial_{\xi}v=\partial_{\xi}w=0$ for deflagrations and one only has  to solve the matching conditions at the shock.

We see that the large enthalpy jump description (solid black) already gives a very accurate estimate of the
wall velocity, even though $N$ is only approximately 3. We also compare the result to the wall velocity obtained assuming local thermal equilibrium at the bubble wall (dashed cyan)~\cite{BarrosoMancha:2020fay, Ai:2021kak, Ai:2023see} and interestingly this gives a comparable estimate, albeit somewhat larger.
This can be understood in the large enthalpy jump limit as follows.
The entropy change at the wall, in the wall frame, is given by
\begin{equation}\label{eq:LTE}
s_H(T_+)\gamma_+v_+ - s_L(T_-)\gamma_-v_- \rightarrow 0,
\end{equation}
which vanishes due to $v_+\rightarrow0$ and $s_-\rightarrow0$ when $N\rightarrow\infty$.
This is exactly the condition for local thermal equilibrium and we would therefore expect an even better agreement between the two approaches for larger $N$. 

Additionally, we display the prediction of the simple wave (sw) formula~\citep{Janik:2022wsx} (pink dot-dashed), with the additional assumption $v_+=0$ and $T_+=T_c$,

\begin{equation}
\xi_w = \tanh\int_{T_n}^{T_c}\frac{dT}{T c_s}.
\label{eq:simple-wave}
\end{equation}
Notice that the results are very close to ours for $T_n \sim T_c$, but start disagreeing for smaller $T_n$. This effect can be perfectly understood in the template model, where both approaches offer analytical results,

\begin{equation}
\begin{aligned}
\xi_w^{{\rm large \, }N} & = \frac{c_{s,H}(T_c^{\mu}-T_n^{\mu})}{\sqrt{(T_c^{\mu}+c_{s,H}^2T_n^{\mu})(T_n^{\mu}+c_{s,H}^2T_c^{\mu})}},\\
\xi_w^{\rm sw} & = \tanh\left(\frac{1}{c_{s,H}}\log\frac{T_c}{T_n}\right),
\end{aligned}
\label{eq:comparison-planar-xi_w}
\end{equation}
with $\mu=1+1/c_{s,H}^2$. The series expansions of both expressions around $T_n = T_c$ agree perfectly to second order. However, greater discrepancy with the data is expected for theories with stronger supercooling.

\section{Implications for GWs}
Let us now discuss the possible implications for the GW spectrum, taking the dark SU(N) model and the 
template model as concrete examples. For the prediction of the GW spectrum, we first follow the approach of~\cite{Caprini:2019egz}, and then further discuss its applicability. 

\begin{figure}[t!]
\begin{center}
\includegraphics[width = 0.45\linewidth]{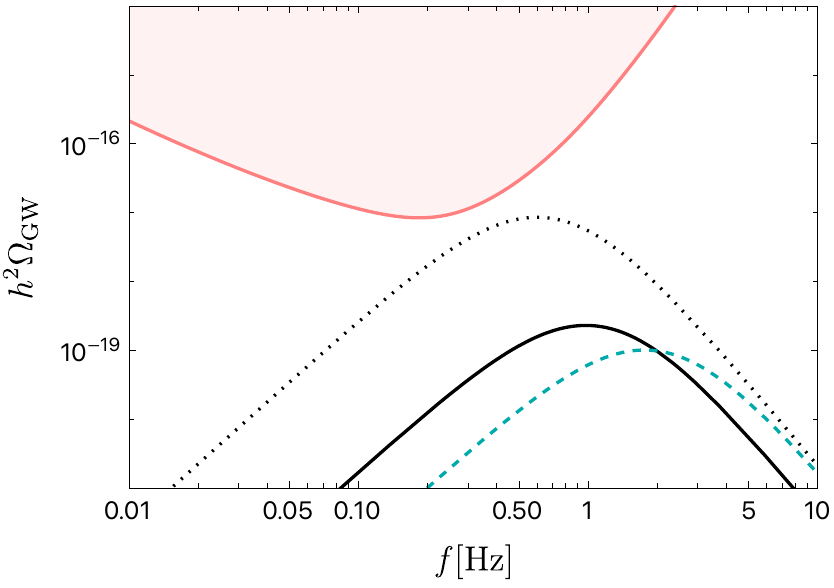}
\caption{Prediction of the GW signal in models with a large enthalpy jump and $T_c = 100 {\rm GeV}$. The two lower lines show the dark SU(N), 
with $T_n/T_c = 0.993$ and $\xi_w$ computed in the large-N limit. The dashed cyan line considers the full model, the black solid line assumed that $c_s^2 = 1/3$. The black dotted line shows the spectrum if one assumes $\xi_w = 1$ and $c_s^2 = 1/3$.
The pink area shows the power-law-integrated sensitivity curve of the future GW experiment BBO~\cite{Crowder:2005nr, Corbin:2005ny, Harry:2006fi}, obtained from~\cite{Schmitz:2020syl}.
}
\label{fig:GWs}
\end{center}
\end{figure}

In~\cite{Morgante:2022zvc}, the GW spectrum of the dark SU(N) model was predicted for three different values of the wall 
velocity $\xi_w = (0.01,0.1,1)$, for a nucleation temperature of $T_n/T_c \sim 0.993 $ (note that $T_n \sim T_p$ holds in this model).
Using the given nucleation temperature,
we find that the correct wall velocity is $\xi_w = 0.078$ in the large enthalpy limit (which is implicitly used in~\cite{Morgante:2022zvc}, by assuming that the low-enthalpy phase can be ignored).
The sound speed deviates strongly from $c_s^2= 1/3$ (see fig.~\ref{fig:high_eos})
and we take its full effect on the energy budget  into account in the dashed cyan line in fig.~\ref{fig:GWs}, 
using $\beta/H = 6.4 \times 10^4 $ and $T_c = 100 \,{\rm GeV}$ from~\cite{Morgante:2022zvc}. For comparison, 
we also show in solid black the GW prediction for $c_s^2 = 1/3$ like was done in~\cite{Morgante:2022zvc}.
We see that the effect of the non-conformal sound speed is to increase the peak frequency by almost a factor 2, 
and to reduce the peak amplitude by a factor $\sim 2$ (largely due to the explicit $c_s$ in the GW amplitude).
If we now compare to the GW predictions of~\cite{Morgante:2022zvc}, we conclude that the large enthalpy jump 
fixes the velocity to a relatively small value, reducing the GW signal compared to more optimistic choices.
Moreover we see that the deviation in $c_s$ significantly affects the signal.
For concreteness, we compared our result with $\xi_w = 0.078$ and the full temperature-dependent $c_s$ and find a suppression of a factor $\sim 80$ compared to the $\xi_w = 1$, $c_s^2 = 1/3$
result obtained in~\cite{Morgante:2022zvc}, demonstrated by the black dotted line.

Although it looks like the GW spectrum could be significantly enhanced by allowing for a larger amount of supercooling, and correspondingly larger $\xi_w$, we refrain from giving such a prediction here. The reason is that the simulations of~\cite{Hindmarsh:2015qta, Hindmarsh:2017gnf} on which the GW predictions of~\cite{Caprini:2019egz} are based never probed large values of $\alpha$ corresponding to such supercooling. Moreover, the vacuum energy domination would trigger a period of inflation \cite{Hambye:2018qjv, Baldes:2018emh}, which affects the GW spectrum.

The GW prediction for models with a large enthalpy jump might even need further revision, due to the large amount of latent heat $L = e_H - e_L \sim e_H$ released. As discussed in \cite{Witten:1984rs, Enqvist:1991xw, Alcock:1987tx, Asadi:2021pwo}, for phase transitions with large amount of latent heat \emph{and} a small amount of supercooling, the nucleation and growth of bubbles gets delayed due to heating of the plasma by the first generation of nucleated bubbles, suppressing the GW amplitude.\footnote{In \cite{Gouttenoire:2023roe}, $\xi_w < 10^{-6}$ was found in this scenario, but note that the amount of supercooling was significantly smaller than the one found in \cite{Morgante:2022zvc}.}
The extent of this effect is possibly model-dependent, as a large amount of latent heat does not need to imply a small amount of supercooling. In \cite{Ares:2021ntv, Bea:2021zol} it was demonstrated that the amount of supercooling in (holographic) models with a large number of degrees of freedom can in fact be large.

Another effect that likely modifies the GW spectrum was observed in the simulations of~\cite{Cutting:2019zws}. For strong deflagrations, bubble walls were observed to slow down before colliding, due to the formation of heated droplets of high-temperature phase. This caused a vortical component in the fluid field, and a suppression of the kinetic energy compared to the predictions of \cite{Caprini:2019egz}. Ref.~\cite{Ares:2020lbt} included this suppression in the prediction of the GW spectrum from a holographic model, and concluded that the spectrum from parameter points with small $\xi_w$ became largely unobservable (by TianQin). Whether the suppression persists for larger phase transition strengths and/or fast hybrid walls is a matter that requires further study.

\section{Conclusion}
In this work we have explored the consequences of a large enthalpy jump, limited allowed temperature range and strong deviations from $c_s^2 = 1/3$ on the wall velocity and the predicted GW spectrum. Although these features arise naturally in holographic and strongly coupled theories, our results are also applicable to weakly coupled theories. 

A large enthalpy-jump highly constrains the fluid flow, favoring deflagrations and hybrids over detonations, which cease to exist as the jump grows larger. It forces the fluid ahead of the wall to be at the critical temperature $T_+=T_c$ and to move at the same speed as the wall, $v_+=0$. By integrating the hydrodynamic equations \eqref{eq:self_similar} and solving matching conditions at the shock we showed that one can obtain a relation between the wall speed and the nucleation temperature depending solely on the EoS. Our estimate for the wall velocity becomes more accurate as the enthalpy jump increases in a given theory, as demonstrated in fig.~\ref{fig:large-N-speed}.
The method can be applied to bubbles in arbitrary dimensions. 
A formula to obtain $\xi_w$ was also proposed in~\cite{Janik:2022wsx}, but it has not yet been generalized beyond planar bubbles, and it only agrees with our result in the regime of small supercooling.

As discussed around eq.~\eqref{eq:LTE}, the large $N$ limit also enforces local thermal equilibrium. This implies that as long as the enthalpy jump is large, the estimate of $\xi_w$ by the code snippet of~\cite{Ai:2023see} (developed for local thermal equilibrium) will be very similar to the one obtained with the method presented in this work.

We have pointed out that a limit on the amount of supercooling, rather than strong coupling, is the main limiting factor in obtaining fast walls. This clarifies the reason behind the low speeds measured in holographic simulations~\cite{Bea:2021,Bea:2022, Janik:2022wsx}.
For illustration, we have computed the GW spectrum in the dark SU(N) model. We found that the limited temperature range indeed results in a small wall speed, which suppressed the GW spectrum compared to the choice $\xi_w =1$. We also demonstrated that non-conformal values of $c_s$, which naturally occur in holographic models, significantly affect the GW spectrum. 

Our main conclusion is that care is needed when determining the wall speed in the computation of the GW spectrum. We have demonstrated a way of estimating the wall velocity in the case of a large enthalpy jump which can even be applied to strongly coupled theories and which should decrease the uncertainty in the GW prediction associated to $\xi_w$.

\section*{Acknowledgements}

We thank Fëanor Reuben Ares, Yann Gouttenoire, Oscar Henriksson, Mark Hindmarsh, Carlos Hoyos, Niko Jokela, Benoit Laurent, Francesco Nitti and Nicklas Ramberg for discussions. We are grateful to Yago Bea for extensive discussions and to Romuald Janik for clarifications on the simple wave formula. %
The work of MSG is supported by the European Research Council (ERC) under the European Union's Horizon 2020 research and innovation program (grant agreement No758759).
JvdV is supported by the Dutch Research Council (NWO), under project  number VI.Veni.212.133.

\newpage
\bibliographystyle{utphys.bst}
\bibliography{refs.bib}

\end{document}